\documentclass[aps,twocolumn,floats,prd,nofootinbib,superscriptaddress,10pt]{revtex4-1}

\usepackage{graphicx}
\usepackage{hyperref}      
\usepackage[usenames,dvipsnames]{color} 
\usepackage{amssymb}
\usepackage{amsmath}
\usepackage{subfigure}

\begin{document}

\title{Constraining Primordial Black Hole Dark Matter with CHIME Fast Radio Bursts }

\author{Keren Krochek}
\affiliation{Department of Physics, Ben-Gurion University Be'er Sheva 84105, Israel}
\author{Ely D. Kovetz}
\affiliation{Department of Physics, Ben-Gurion University Be'er Sheva 84105, Israel}

\date{\today}

\begin{abstract}
Strong lensing of Fast Radio Bursts (FBRs) has been proposed as a relatively clean probe of primordial black hole (PBH) dark matter. Recently, the Canadian Hydrogen Intensity Mapping Experiment (CHIME)  published a first catalog of 536 FRBs, 62 of which are from repeating sources. In light of this new data, we re-examine the prospects to constrain the abundance of PBHs via FRBs. Extending previous forecasts, we calculate a PBH dark matter bound using the intrinsic burst width and a calibrated flux-ratio threshold per FRB. 
In addition, we take into account the uncertainty in the relation between the FRB dispersion measure and source redshift.  We outline an algorithm to detect lensed FRBs and a method to simulate its performance on real data and set a flux-ratio threshold for each event, which we use to infer realistic forecasts. We then attempt to extract a preliminary bound using the publicly available CHIME data. Unfortunately, both instrumental noise and the  provided $\sim\!1\,{\rm ms}$ time-resolution of the public data hinder this effort. We identify one candidate event where a double burst could be explained via strong lensing by a $\mathcal{O}(10\,M_\odot)$-mass PBH, which will require follow-up study at higher time resolution to either confirm or discard. We show that with a few times the size of the first catalog---sampled at the full instrumental time-resolution so that candidates can be efficiently scrutinized---CHIME will be able to find strong evidence for or robustly rule out PBHs with mass above $\sim\!10\, M_\odot$ as the dark matter. Finally, we demonstrate that stacking repeating FRBs can improve the constraints, especially for lower  masses.
\end{abstract}

\maketitle

\section{Introduction}\label{sec:intro}

Primordial black holes (PBHs)  formed early in the radiation-dominated era are an intriguing explanation for the dark matter in the Universe (for recent reviews, see Refs.~\cite{Sasaki:2018dmp,Green:2020jor}). In particular, if their masses are in the $10\,M_\odot \lesssim M\lesssim 100\,M_\odot$  range, they could potentially account for some of the detected gravitational-wave merger events in the Laser Interferometer Gravitational-Wave Observatory~\cite{LIGOScientific:2007fwp} (LIGO) and the Virgo interferometer \cite{VIRGO}, see for example Refs.~\cite{Bird:2016dcv,Clesse:2016vqa,Sasaki:2016jop,DeLuca:2020qqa,Jedamzik:2020omx,Wong:2020yig,Franciolini:2021tla}.

Multiple ways of looking for PBHs in this mass range have been considered~\cite{Carr:2020gox}. One promising method, which draws inspiration from the microlensing method used to search for massive compact halo objects (MACHOs) for several decades now, is to look for a signature of strong-lensing of fast radio bursts (FRBs) due to intervening PBHs along the line of sight~\cite{Munoz:2016tmg}. FRBs, which are transients of millisecond duration and  extragalactic origin, are an excellent probe of PBH dark matter in the LIGO mass range, as the optical depth for strong lensing would be large enough to yield some detections given the large number of FRBs~\cite{Fialkov:2017qoz} and as the separation between the original and lensed images would be of the same order as the intrinsic burst width, making them easy to detect.

In this paper we set out to constrain the fraction of dark matter in PBHs with the new CHIME FRB catalog~\cite{CHIMEFRB:2021srp}, using a dedicated algorithm to set an individual flux-ratio detectability threshold per FRB, and accounting for the uncertainty in the dispersion measure to redshift relation~\cite{James:2021jbo}.  
We find that current data is insufficient to provide a meaningful bound.  However, we identify one FRB with an echo that could in principle be explained via lensing due to $\mathcal{O}(10\,M_\odot)$-mass PBH and demonstrate that with proper validation of the dataset, a robust bound can be placed on PBH dark matter with a few times the size of the first CHIME catalog. We also argue that stacking of repeating FRBs can potentially be used to improve the bound and probe lower PBH masses.

\section{Data}

We use the recently released 1st catalog of CHIME~\cite{CHIMEFRB:2021srp}. The catalog contains a total of 536 FRBs, 62 of which originate from 18 repeating sources. Below we describe the signal-to-noise, width, inferred redshift and repeater-burst distributions of the CHIME FRBs. These will enable, using the methodology outlined in Section III, to extract a bound on the fraction of dark matter in PBHs.

\subsection{Redshift, width and signal-to-noise distributions}

FRBs have accurately measured dispersion measures (DMs),  resulting from interaction with free electrons along the line of sight to their source~\cite{Petroff:2019tty,Cordes:2019cmq,Petroff:2021wug}. 
Therefore, with a model for the free electron distribution in each FRB path, we can infer its redshift from the directly measured DM~\cite{Kumar:2019qhc}. We  model the  DM as consisting of 
\begin{equation}\label{eq:DM}
{\rm DM}={\rm DM}_{\rm MW}+{\rm DM}_{\rm EG},
\end{equation}
where
\begin{equation}
{\rm DM}_{\rm EG}={\rm DM}_{\rm cosmic}+\frac{{\rm DM}_{\rm host}+{\rm DM}_{\rm src}}{1+z}
\end{equation}
is the extra-galactic DM contribution, ${\rm DM}_{\rm MW}$ is the total Milky-Way contribution, ${\rm DM}_{\rm cosmic}$ is the contribution of the intergalactic medium (IGM), and ${\rm DM}_{\rm host}$ and ${\rm DM}_{\rm src}$ result from the FRB host galaxy the its source environment, respectively (and are hence divided by $(1+z)$). 

We  use the NE2001 model~\cite{Cordes:2002wz} to calculate ${\rm DM}_{\rm MW}$. For the extra-galactic contribution, we follow  Ref.~\cite{James:2021jbo} and adopt their $z-{\rm DM}_{\rm cosmic}$ relation between redshift and DM, where an upper bound is calibrated for the combination of the FRB source environment and host galaxy (we will use their best-fit value ${\rm DM}_{\rm host}+{\rm DM}_{\rm src}=145\,{\rm pc}\, {\rm cm}^{-3}$). For the MW halo, we use ${\rm DM}_{\rm halo}=50$.

As for the burst widths, there are two types of widths for each FRB in the catalog. The first is the intrinsic width (denoted as {\tt width\_fitb} in the CHIME catalog). These widths are obtained using the {\tt  findburst} function, which has not been made public, and so we take them directly from the catalog. 
The second is the boxcar width ({\tt bc\_width} in the catalog). These widths are multiples of $0.98304\,{\rm ms}$ and are not optimal proxies for the intrinsic burst widths, however this is the resolution of the publicly available Waterfall data files~\cite{Waterfall} and therefore these are the widths used in our detection algorithm{\footnote{If they are made available, a more precise analysis can be done using the CHIME baseband data that are saved to disk for at least some of the CHIME FRBs, with a time resolution  of $2.56\mu s$~\cite{Pleunis:2021qow}.}${}^,$\footnote{Some complex FRBs are comprised of multiple sub-bursts \cite{Pleunis:2021qow}. For these the boxcar width encapsulates multiple sub-bursts. The intrinsic widths of the sub-bursts vary and some are shorter than the time resolution, making them undetectable through autocorrelation. These bursts make up roughly $10\%$ of the FRBs. Each sub-burst in the Catalog is assigned a sub-number which denotes its chronological order. The first sub-burst and FRBs without sub-bursts are assigned with ``$0$", and these are the bursts we use in our analysis. Since we only probe for lensing with a minimal full-width separation between the original and lensed images, we do not detect the sub-bursts as lensed candidates.}}.  

\begin{figure}[htbp!]
	\includegraphics[width=0.43\textwidth]{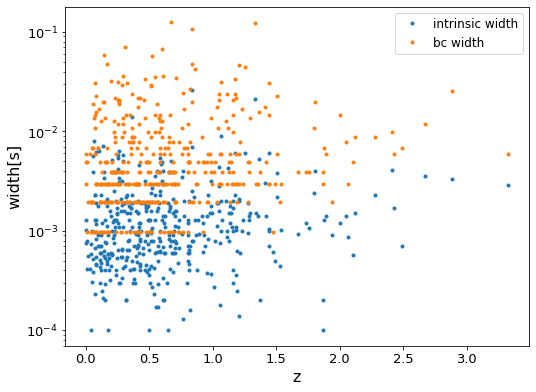}
	\includegraphics[width=0.43\textwidth]{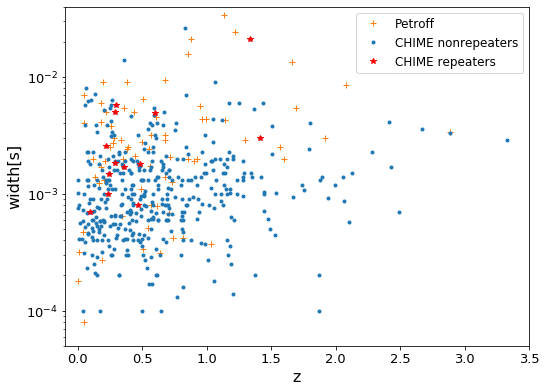}
	\caption{{\it Top}:  boxcar widths (orange) and intrinsic widths (blue) vs.\  $z$.  {\it Bottom}: intrinsic width vs.\ $z$ for CHIME FRBs (blue dots for non-repeaters, red stars for repeaters) and for FRBCat FRBs~\cite{Petroff:2016tcr} (orange $+$).}
	\label{fig:z-widths}
\end{figure}

In Fig.~\ref{fig:z-widths}, we show the boxcar {\tt bc\_width} and intrinsic {\tt width\_fitb} burst widths as a function of the inferred FRB redshift $z$ (note that the former do not reach values smaller than $\sim\!1\,{\rm ms}$, which are critical to probing lower PBH masses). 
In the top panel of Fig.~\ref{fig:z-widths}, we show the intrinsic widths of Catalog 1 (distinguishing repeaters from non-repeaters) as well as those from the former FRBcat catalog~\cite{Petroff:2016tcr}\footnote{We omit several FRBs from the Pushchino telescope which have abnormally large widths.} vs.\ redshift $z$. In both cases we use  the $z-{\rm DM}_{\rm cosmic}$ prescription to convert DM to $z$. For some nearby FRBs this calculation yields a negative $z$. As these do not contribute significantly to our PBH constraint, we simply discard them. From now on we denote the number of remaining FRBs as $N_{\rm FRB}$.

The measured ${\rm DM}$ and the signal-to-noise ratio (SNR) we use below are also products of the ${\tt findburst}$ function and are taken directly from fields ${\tt dm\_fitb}$ and ${\tt snr\_fitb}$ in the catalog. Note that
${\rm DM}_{\rm EG}$ is the ${\rm DM}$ excess between the total observed ${\rm DM}$ determined by fitburst and the NE2001 model, using the best-fit sky position of the source, which is given in the catalog as ${\tt dm\_exc\_ne2001}$.

\begin{figure}[htbp!]
	\includegraphics[width=0.43\textwidth]{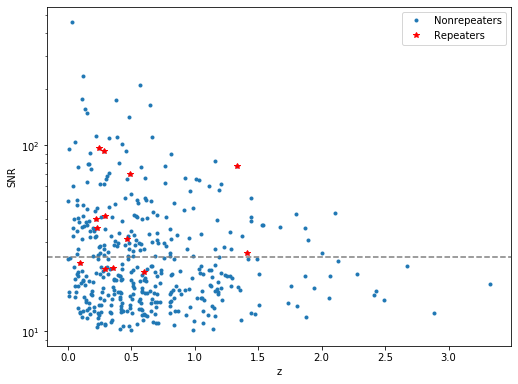}
	\caption{SNR vs.\ redshift $z$. Dashed line marks ${\rm SNR}=25$; we use red stars to indicate repeaters and blue dots  nonrepeaters.}
	\label{fig:z-snr}
\end{figure}

In Fig \ref{fig:z-snr} we show the SNRs of the FRBs as a function of their redshift $z$. We see that high SNRs become less common at high redshift. For example, out of 536 CHIME FRBs in total, there are $232$  FRBs with ${\rm SNR}\geq25$.

\subsection{Repeater statistics}

\begin{figure}[htbp!]
	\includegraphics[width=0.23\textwidth]{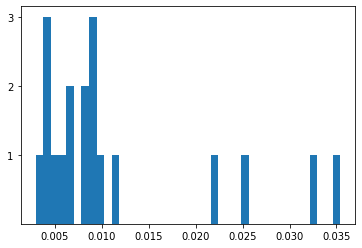} 
    \includegraphics[width=0.23\textwidth]{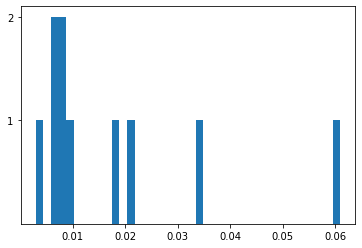}
	\caption{Histograms of boxcar widths (in secs) of burst repetitions, for FRB20180814A (left) and FRB20180916B (right).}
	\label{fig:widths hist}
\end{figure}

A subset of the total FRB population is known to repeat (the repeaters may belong to a different channel, see e.g.\ Ref.~\cite{Pleunis:2021qow}). The CHIME catalog includes 18 repeaters, each with a different number of repetitions.  As we shall advocate, stacking these FRB repetitions could increase the sensitivity to lensing and thereby improve the extracted bound on dark matter in the form of PBHs. However, for stacking to work, the repetitions must not vary wildly in width. 
In Fig.~\ref{fig:widths hist} we show the histograms of boxcar widths for repeating FRBs FRB20180916B and FRB20180814A, as examples of FRBs in the current catalog with more than 10 repetitions. While some of the repetitions are considerably wider, both FRBs have a $\gtrsim10$ sub-group of bursts with widths $\lesssim\!10\,{\rm ms}$ which may be useful in a stacking analysis, as we describe below.  

\section{Method}\label{sec:method}

\subsection{Deriving a constraint on PBH dark-matter  from non-detection of strong lensing of FRBs}
Below we derive a bound on PBH dark matter based on the combined optical depth for a population of FRBs with different intrinsic burst widths and thresholds for detectable flux ratios between original and lensed images.

In general, a MACHO of mass $M_L$ can be treated as a point lens with an (angular) Einstein radius
\begin{equation}
\theta_{E}=2\sqrt{\frac{GM_L}{c^2}\frac{D_{LS}}{D_SD_L}},
\end{equation}
where $D_S$, $D_L$ and $D_{LS}$ are the angular diameter distances to the source, to the lens, and between the source and the lens, respectively. A point lens produces two images, at positions $\theta_{\pm}=(\beta\pm\sqrt{\beta^2+4\theta_E^2})/2$, where $\beta$ is the angular impact parameter. The time delay between these two images is given by
\begin{equation}
\bigtriangleup t=\frac{4GM_L}{c^3}(1+z_L)\left[\frac{y}{2}\sqrt{y^2+4}+\log{\left(\frac{\sqrt{y^2+4}+y}{\sqrt{y^2+4}-y}\right)}\right],\\
\label{eq:timedelay}
\end{equation}
where $y \equiv\beta/\theta_E$ is the normalized impact parameter and
$z_L$ is the redshift of the lens. We also define the 
flux ratio $R_f$ as the absolute value of the ratio of the magnifications $\mu_+$ and $\mu_-$
of both images as
\begin{equation}
R_f \equiv \left|\frac{\mu_+}{\mu_-}\right|=\frac{y^2+2+y\sqrt{y^2+4}}{y^2+2-y\sqrt{y^2+4}}>1.
\end{equation}

Note that for a given lens mass $M_L$ and lens redshift $z_L$, both $R_f$ and $\Delta t$ are monotonic functions of the impact parameter $y$. For the lensing-induced echo to be detectable, the impact parameter must satisfy two conditions. First, the observed time delay must be larger than some reference time $\overline{\bigtriangleup t}$, which will place a lower bound on the impact parameter $y\!>\!y_{\rm min}(M_L,z_L,\overline{\bigtriangleup t})$.
Secondly, there exists a maximal detectable magnification ratio $\overline{R}_f$, where for any $R_f > \overline{R}_f$ the second image is too weak to be detected, which then places an upper bound on the impact parameter $y<y_{\rm max}(\overline{R}_f)$. Together, the two limits determine an effective cross section
\begin{equation}
\sigma(M_L,z_L)=\pi\theta^2_E D_L^2 \left[y_{\rm max}^2({\overline{R}_f})-y_{\rm min}^2(M_L,z_L,\overline{\bigtriangleup t})\right],\\
\end{equation}
with the dependence on $\overline{R}_f$ and $\overline{\bigtriangleup t}$ henceforth implicit.

This cross section can be used to calculate the probability for an FRB to be lensed. 
The lensing optical depth of a source at redshift $z_S$ is given by
\begin{equation}
\tau(M_L,z_S)=\int_{0}^{z_S}d\chi(z_L)(1+z_L)^2n_L\sigma(M_L,z_L),
\label{eq:tau4}
\end{equation}
where $\chi(z)$ is the comoving distance at redshift $z$ and $n_L$ is the comoving number density of lenses.
In order to calculate the integrated lensing probability, the optical depth for lensing of a single burst has to be convolved with the redshift distribution of incoming FRBs.
Given a distribution function $N(z)$ for FRBs, we can calculate their integrated strong-lensing optical 
depth $\bar \tau(M_L)$, due to a monochromatic population of PBHs of mass $M_L$, as
\begin{equation}
\bar \tau (M_L) = \int dz\, \tau(z,M_L) N(z).
\label{eq:tau_bar}
\end{equation}
Since we are in the optically-thin regime,  the probability to be lensed is just $P_{\rm lens} = 1-e^{-\bar \tau}\approx \bar \tau$. Thus, if we observe a number $N_{\rm FRB}$ of FRBs, $\bar \tau  N_{\rm FRB}$ of them should be lensed.
So, for example, if we do not detect any lensed events, we can infer a PBH constraint $f_{\rm DM}<{1}/{\bar\tau N_{\rm FRB}}$, where $f_{\rm DM}$ is the fraction of total dark matter in PBHs.

Rather than assuming (or fitting for) a redshift distribution $N(z)$, when working with a catalog of real events we will replace the integral in Eq.~\eqref{eq:tau_bar} with a simple sum
\begin{equation}
\bar \tau' (M_L) = \sum_{i} \tau(z_i,M_L,w_i,\overline{R}_{f,i}),
\label{eq:tau_bar_sum}
\end{equation}
where $z_i$ and $w_i$ are the redshift and width of each FRB, and we explain below how $\overline{R}_{f,i}$ is set for each FRB. The optical depth $\tau$ per FRB is a function of its width, as we take $\overline{\bigtriangleup t}_i = w_i$ for the calculation of $y_{\rm min}$, and the sum is over all FRBs.
Finally, we extract the bound $f_{\rm DM}<{1}/{\bar \tau'}$.

We shall see below that in practice, the full CHIME catalog cannot be used in search for lensing echos as many FRBs have signal-to-noise ratios (SNRs) too low to allow for detection and they must be filtered. In Section~\ref{sec:results}  we attempt to derive a bound from the first catalog, and provide data-based realistic forecasts for future releases.

\subsection{Accounting for uncertainty in redshift-DM relation}

Previous studies~\cite{Munoz:2016tmg,Liao:2020wae} have ignored the uncertainty in the derived bound on dark matter stemming from the fiducial DM to redshift relation they used.
Here we compare the effect of two prescriptions for this relation on the resulting uncertainty in the derived $f_{\rm DM}$ constraint. 

\begin{figure}[htbp!]
	\includegraphics[width=0.43\textwidth]{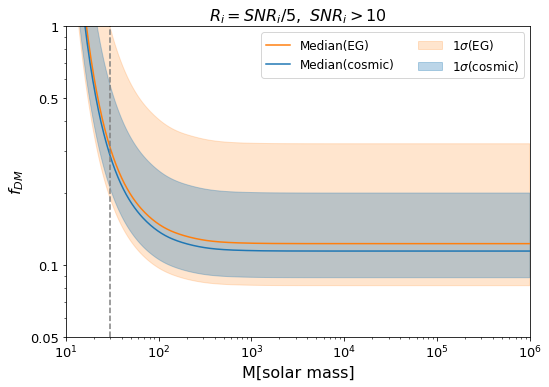}
	\caption{Fraction $f_{\rm DM}$ of dark matter allowed in the form of PBHs of mass $M$, if no events out of $N_{\rm FRB}$ are lensed, using the $z-{\rm DM}_{\rm cosmic}$ (blue) and $z-\rm DM_{EG}$ (orange) relations.}
	\label{fig:fDM cosmic vs EG}
\end{figure}

One is the $z-{\rm DM}_{\rm cosmic}$ relation described in the previous section. The other is the $z-{\rm DM}_{\rm EG}$, also taken from Ref.~\cite{James:2021jbo}, which makes no assumptions about the source environment and yields more conservative uncertainty ranges. 
In Fig.~\ref{fig:fDM cosmic vs EG} we show a forecast for the bound on $f_{\rm DM}$ that can be placed with CHIME FRBs setting a universal ${\rm SNR}/5$ flux-ratio threshold (and a corresponding minimum ${\rm SNR}\geq10$ cuttoff), along with the uncertainty due to the $z-{\rm DM}$ relations. In the rest of the paper, we use the more informed $z-{\rm DM}_{\rm cosmic}$ relation.

%For this example we make the same assumptions as described in the forecast section below.
%Using the second $z-{\rm DM}$ distribution from Ref.~\cite{James:2021jbo}, which makes no assumptions about the source environment, yields more conservative uncertainty ranges.

\subsection{Strong lensing detection algorithm}

In this subsection we describe our method to search for strongly lensed FRBs. We assume that the location and width of the burst itself are already identified in the data. In application to the CHIME catalog, we use the ${\tt cfod}$ python package~\cite{CHIME utilities} and the CHIME FRB Waterfall Data (see Ref.~\cite{Waterfall}).
Recall that the available data is sampled in intervals of $dt\!\sim\!1\,{\rm ms}$ and hence we can only use the boxcar widths  ${\tt bc\_widths}$. To isolate the burst, we cut out a section with length ${\tt bc\_widths}$ to either side of its peak. 
This section is treated as the {\it burst} and the rest of the signal is treated as the {\it noise} (see an example of an identified burst in Fig.~\ref{fig:detection example}; the burst is shaded in orange).

We then follow the algorithm proposed in  Ref.~\cite{Ji:2018rvg}, where autocorrelation was used to detect strong gravitational lensing of gamma ray bursts (GRBs) by massive compact halo objects (or PBHs) in order to try to constrain $f_{\rm DM}$ in the mass range $10\,M_\odot\! \lesssim\! M \!\lesssim \!1000\,M_\odot$. Unlike GRBs, for which lenses in this mass range do not produce two separate images, but rather a single image of overlaid signals, FRBs will appear as two separate images when strongly lensed due to their short intrinsic widths.

The autocorrelation function of the light curve $I(t)$ is
\begin{equation}
C(\delta t) = \frac{\int dt\, I(t)I(t-\delta t)}{\sqrt{\int dt\, I^2(t)}\sqrt{\int dt\, I^2(t-\delta t)}},
\end{equation}
which is a normalized cross correlation between $I(t)$ and its shifted self. Note that $C(0)\! =\! 1$. For generic non-autocorrelated functions,
$C(\delta t \neq 0)<1$, and thus $\delta t = 0$ serves as the only major spike in the function $C(\delta t)$. We will  require $C(\delta t)$ beyond the initial spike to be greater than some threshold  for an FRB to be considered lensed. 

The ideal way to set the threshold for detection of a lensed image of a given burst would be to isolate only the burst signal itself along with a {\it clean} stretch of pure instrumental noise. However, for any burst in the catalog, the post-burst signal may contain a lensed echo. Therefore,  we first  take the pre-burst signal (which is guaranteed not to contain a lensed repetition) along with the burst itself and calculate the autocorrelation (we first flip this section to place the burst at the beginning, see Fig.~\ref{fig:detection example}).
We then calculate the standard deviation $\sigma$ of this autocorrelation after cutting the initial peak which is due to the burst overlapping with itself (see the autocorrelation plot with the gray shaded region in Fig.~\ref{fig:detection example}). 

Finally, we calculate the autocorrelation of the full FRB time series and set the threshold to be $2.325\sigma+\mu$ (corresponding to a detection of an outlier at $99\%$-C.L.), where $\mu$ is the mean of the new autocorrelation (see blue horizontal line in Fig.~\ref{fig:detection example}). We then look for any peaks higher than this threshold in the region to the right of the end of the first peak (see dashed vertical line in Fig.~\ref{fig:detection example}, which marks the first uptick in the autocorrelation). If we find such a peak, we have a candidate lensed FRB(!).

\subsection{Algorithm validation via data-based simulations}\label{subsec:data validation}
 
Many FRBs have SNRs too low to be detected as lensed, so before claiming an FRB as unlensed we first check for which $R_f$ it can be detected as lensed by our algorithm, if any. We do this by running the algorithm on simulated lensed signals with different  flux ratios $R_f$.

\begin{figure}[h!]
	\includegraphics[width=0.23\textwidth]{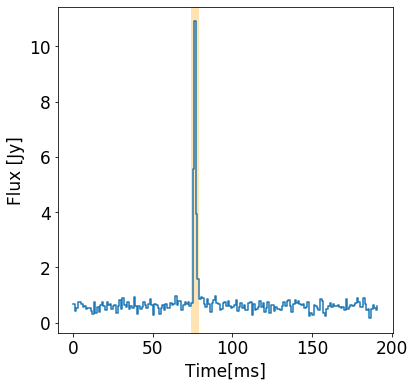}
	\includegraphics[width=0.23\textwidth]{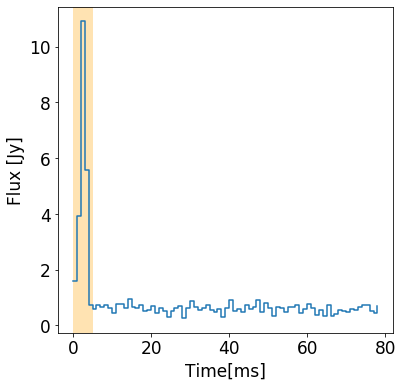}
	\includegraphics[width=0.23\textwidth]{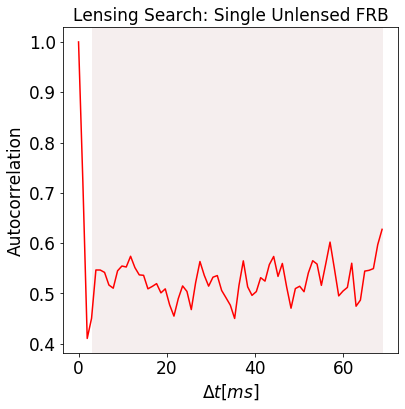}
	\includegraphics[width=0.23\textwidth]{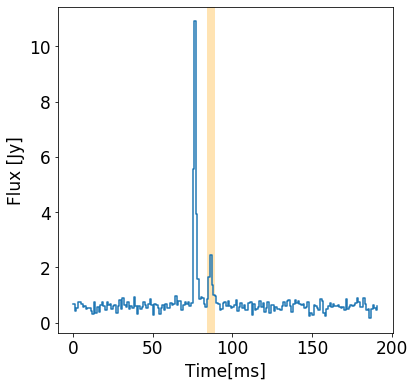}
	\includegraphics[width=0.25\textwidth]{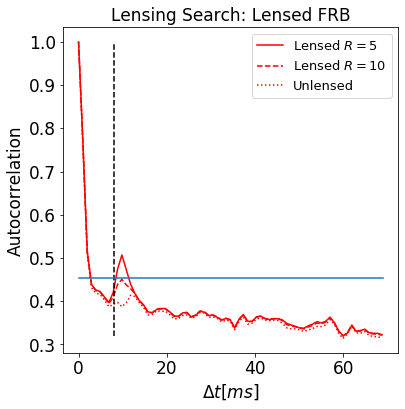}
	\caption{Example: validating our detection algorithm against FRB20190118A.
	{\it Upper left:} Full time series with the burst highlighted in orange. {\it Upper right:} Section from the beginning to the end of the burst, flipped to place the burst at the beginning. {\it Middle left:} Autocorrelation of the signal in the upper right panel. The standard deviation $\sigma$ used to set the threshold is calculated for the shaded part. {\it Middle right:} A simulated lensed signal with the echo shaded in orange. {\it Bottom:}  Algorithm results. The dashed line marks the beginning of the search. The $99\%$-C.L.\ detection threshold is shown in solid blue. The red lines are the autocorrelation results, solid for $R_f=5$, dashed for $R_f=10$ and dotted for the original unlensed signal. This FRB passed detection up to $R_f=10$.}
	\label{fig:detection example}
\end{figure}

The simulated lensed signals are created as follows: we take the {\it burst} and subtract from it the mean of the {\it noise}, to avoid adding noise along with the repetition (thus doubly-counting it). We then shift this {\it echo} forward in time with a default shift large enough to ensure full separation between the images (we have checked that results are not sensitive to a particular choice of shift), divide it by $R_f$, and add it to the original signal (see the {\it echo} shaded in orange in Fig.~\ref{fig:detection example}). Note that just like our {\it burst}, the {\it echo} we add is a section twice the ${\tt bc\_widths}$ in length. This is done to make sure we do not miss any part of the burst (we found that many bursts in the catalog extend beyond their {\tt bc\_widths}). We deem this safe as we subtract  the mean of the {\it noise} from the echo, so it does not matter if the echo extends a bit into the {\it noise}.

To determine how suitable an FRB is for lensing detection,  we run the algorithm on its simulated lensed version for different values of $R_f$, while demanding that the place where the autocorrelation crosses the threshold is the same as where we simulated the echo. The maximum value for which a detection is made sets $\overline{R}_{f,i}$ for this FRB. These  $\overline{R}_{f,i}$ are  then used in Eq.~\eqref{eq:tau_bar_sum} to set the overall $f_{\rm DM}$ threshold.

\section{results}\label{sec:results}

Before we attempt to extract or forecast a bound on $f_{\rm DM}$ based on the CHIME catalog, we filter the dataset and implement our validation procedure to remain with an effective group of FRBs to be used in the search for echos from strong lensing by PBHs along the line of sight.

\subsection{Validation of CHIME data}

\begin{figure}[h!]
	\includegraphics[width=0.45\textwidth]{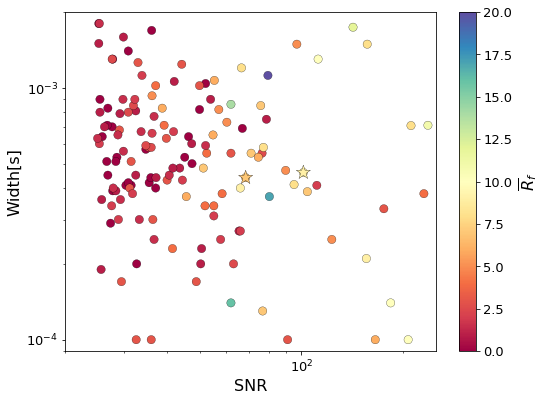}
	\caption{The $w_i$-$\overline{R}^i_f$-${\rm SNR}$ distribution for the filtered 143 CHIME FRBs, 114 of which are validated by our procedure.}
	\label{fig:Rf_per_SNR}
\end{figure}

\begin{figure}[h!]
	\includegraphics[width=0.48\linewidth]{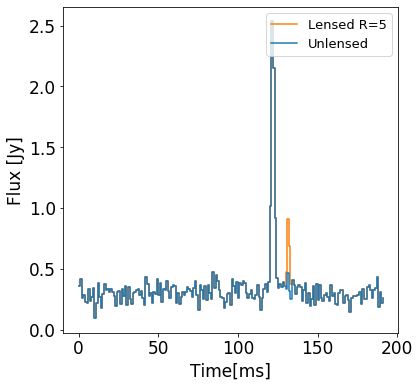}
	\includegraphics[width=0.48\linewidth]{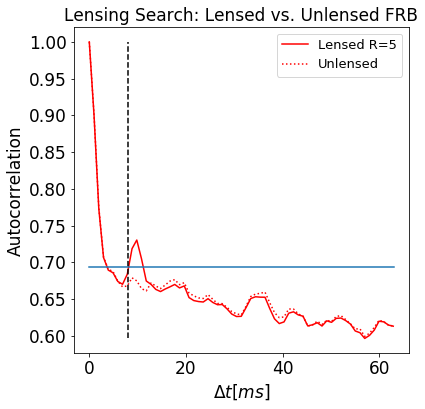}
	\includegraphics[width=0.48\linewidth]{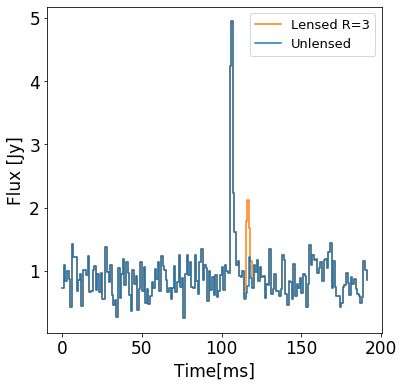}
	\includegraphics[width=0.48\linewidth]{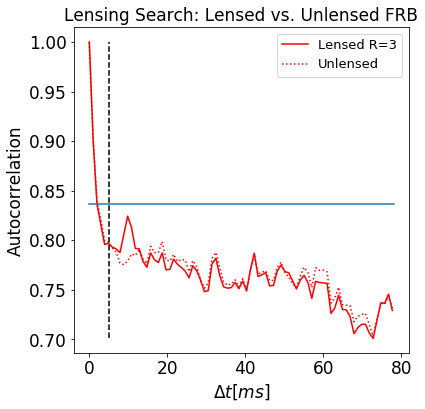}
	\includegraphics[width=0.48\linewidth]{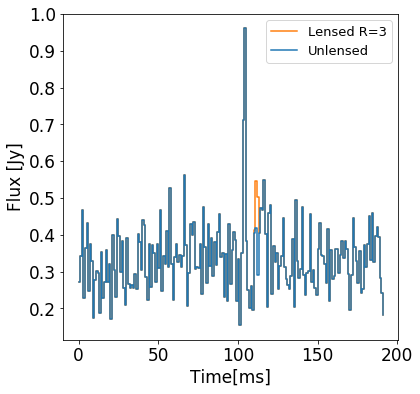}
	\includegraphics[width=0.48\linewidth]{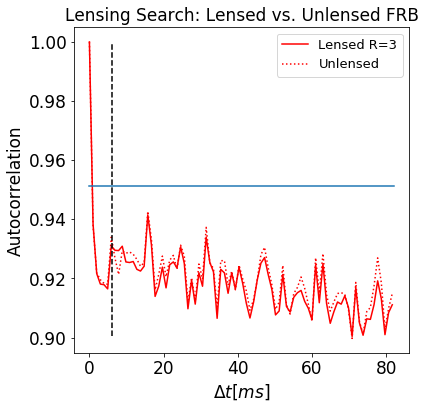}
	\caption{Examples of the data validation process.  Left panels show the time series of the bursts (blue) and the added echo (orange), while right panels show the autocorrelation of the lensed  (solid) and unlensed (dotted) signals.
	{\it Top row:} FRB20190215B with ${\rm SNR}\!=\!61.9$ which is validated with $\overline{R}^i_f\!=\!14$. {\it Middle row}: FRB20190427A, with ${\rm SNR}\!=\!27.5$, but  validation fails even for $R_f\!=\!3$. {\it Bottom row}: FRB20190213C, with ${\rm SNR}\!=\!11.5$, which is also not validated with $R_f\!\geq\!3$.}
	\label{fig:validation example}
\end{figure}

We first make two gross cuts to the CHIME FRB data. We filter out FRBs with widths $\geq 6\,{\rm ms}$---which are too wide to constrain PBHs in the desired mass range\footnote{%\color{red}
We focus on the range $10-100\,M_\odot$ as it is motivated by LIGO~\cite{Bird:2016dcv} and as there already are sufficient constraints above $100\,M_\odot$~\cite{Carr:2020gox}.}---and with ${\rm SNR}\!<\!25$\footnote{This choice is somewhat arbitrary, alternatively we could remove the SNR filtering and in turn set a higher autocorrelation threshold to filter out spurious detections, as we discuss in Section~\ref{sec:conclusions}.}, as the noise in these FRBs is simply too large to allow a detection of a  lensed echo (see Figs.~\ref{fig:Rf_per_SNR},\ref{fig:validation example}). 

This cut leaves us with only 143 FRBs to be validated using the procedure described in the previous section before  being used to derive a bound on $f_{\rm DM}$ via Eq.~\eqref{eq:tau_bar_sum}. 

For each of the 143 FRBs we run a series of simulations and find the maximal $R_f$ in the range $\{1,1.5,2, 3,...,20\}$ that allows a detection of a simulated echo.  Fig.~\ref{fig:validation example}  shows three examples of this validation process, one for an FRB with a high SNR which passes the validation process, one with a medium SNR which passes the initial SNR cut but fails the validation process and one with a low SNR, where the added echo is of the same intensity as the noise.

A  total of 114 FRBs pass the validation process, with a width ($w_i$) to maximal-flux-ratio ($\overline{R}^i_f$) distribution as shown in Fig.~\ref{fig:Rf_per_SNR}. This pristine set of FRBs can  be reliably used to search for lensing-induced echos and bound $f_{\rm DM}$.

\subsection{Detected lensing candidates}

\begin{figure}[h!]
	\includegraphics[width=0.23\textwidth]{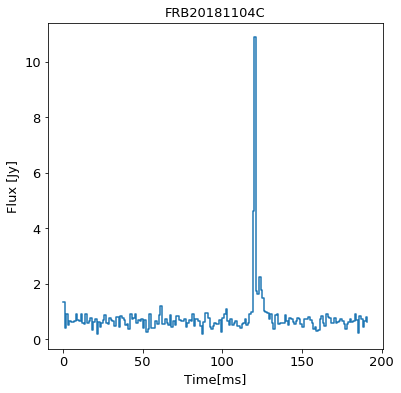}
    \includegraphics[width=0.23\textwidth]{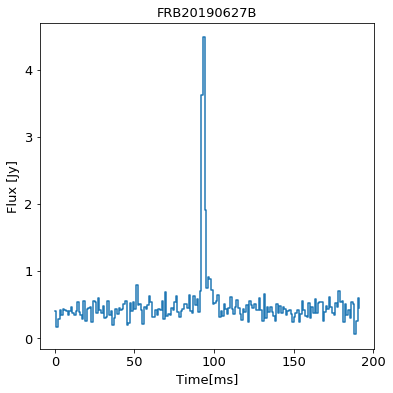}
    \includegraphics[width=0.23\textwidth]{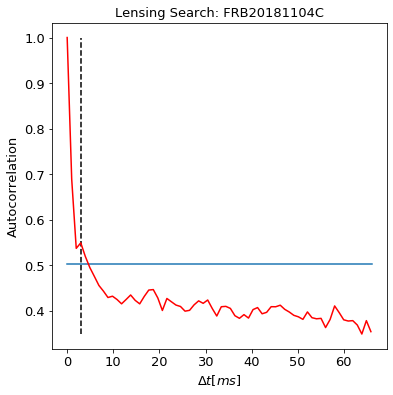}
    \includegraphics[width=0.23\textwidth]{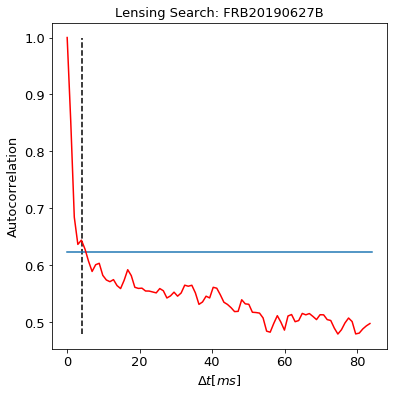}
\vspace{-0.125in}
	\caption{Time series and autocorrelation for our two candidate detections: FRB20181104C (left) and FRB20190627B (right).  For FRB20181104C, the second peak is $\sim\!3$ times wider than the first (as shown in the CHIME catalog, based on higher-resolution data) and thus we can rule it out as a lensed FRB.}
	\label{fig:lensed}
\end{figure}

We now run the search algorithm on the original unlensed  time series of the 114 validated FRBs identified above. Among these, we find three candidate events. Two of them, FRB20190627B and FRB20181104C, have a similar profile---showing an echo at a $\sim\!3\,{\rm ms}$ separation from the main burst---which in principle could be explained by strong lensing due to a $\mathcal{O}(10\,{\rm M_\odot})$ PBH.
Their time series and autocorrelation are shown in Fig.~\ref{fig:lensed}. For FRB20181104C, our search in fact picked up a sub-burst which according to the CHIME catalog has an intrinsic width {\tt width\_fitb} (which is determined using higher-resolution data than  is publicly available) that is $\sim\!3$ times as wide as the main burst, and so it can be eliminated as a lensing event. For FRB20190627B, the catalog does not show a sub-burst detected at higher-resolution.

\begin{figure}[h!]
	\includegraphics[width=0.23\textwidth]{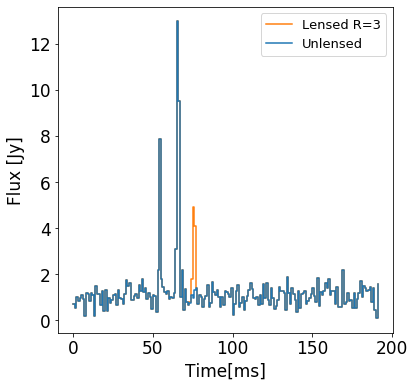}
    \includegraphics[width=0.23\textwidth]{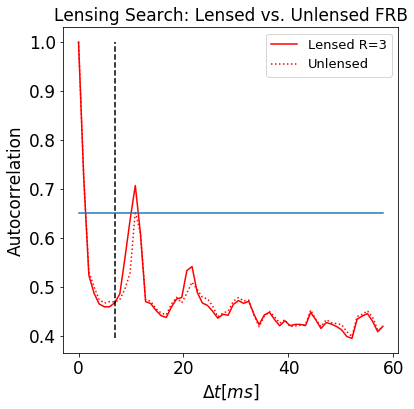}
\vspace{-0.125in}
	\caption{Time series and autocorrelation for FRB20190609B. This FRB passes our validation process with $R=3$, and is then detected as a candidate in our lensing search. However, as the second burst is brighter, this cannot be a lensed FRB.}
	\label{fig:false detection}
\end{figure}

In addition, the third candidate, FRB20190609B, can be immediately ruled out as a lensing candidate as its pre-burst echo is dimmer than the main burst (see Fig.~\ref{fig:false detection}), which is impossible if due to strong lensing (as long as the point-source lens approximation is valid).
%{\color{red}
In Section~\ref{sec:DS} we use the full FRB dynamic spectrum to further scrutinize these candidates using an energy hardness test.

%\vspace{-0.15in}

\subsection{A preliminary (non)-bound from CHIME}

\begin{figure}[h!]
\includegraphics[width=0.45\textwidth]{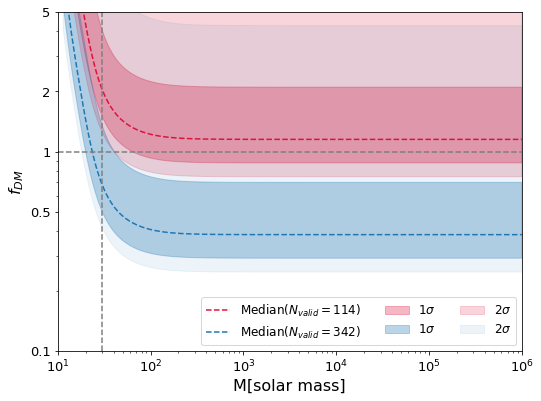}
\caption{Fraction $f_{\rm DM}$ of dark matter allowed in the form of
point lenses of mass $M_L$, if $\leq1$ events out of $N_{\rm valid}=114$ are lensed (dashed red), with {%\color{red}
corresponding $1\sigma$ and $2\sigma$ uncertainties from the $z-{\rm DM}_{\rm cosmic}$ relation ($2\sigma$ upper limit extends to $\sim 13$)}. The dashed blue line shows a bound with thrice the data of the first CHIME catalog. The vertical dashed  line marks $M=30\,M_\odot$, for orientation.\color{black}}
\label{fig:fDM}
\end{figure}

Based on our findings: one lensed ``detection" out of 114 events, with the $\overline{R}^i_f$-$w_i$ distribution shown in Fig.~\ref{fig:Rf_per_SNR}, we can immediately extract a bound using Eq.~\eqref{eq:tau_bar_sum} and $f_{\rm DM}<{1}/{\bar\tau N_{\rm FRB}}$. This ``bound" is shown in Fig.~\ref{fig:fDM} (red). As can be seen, this lies above the abundance required to explain the totality of dark matter at all mass scales (indicating that our candidate may be spurious), but within the uncertainty from the DM-redshift relation.

%\vspace{-0.3in}

\subsection{Data-based forecasts}\label{sec:Forecast}
%\vspace{-0.1in}

With only one lensing candidate remaining, the bound that is inferred on $f_{\rm DM}$ becomes consistent with PBH dark matter (see the dashed red line in Fig.~\ref{fig:fDM} above). 

Looking forward, with more CHIME data expected in the near future, a robust bound can be improved to become meaningful and gradually more significant. 
 Note that with the full time-resolution the validation procedure may yield a higher number of valid FRBs and thus enable a stronger bound even with current data. With just thrice the number of valid FRBs we found in the first CHIME catalog, the bound can fall below $f_{\rm DM}\!=\!0.5$. With another doubling of the data, a value of unity will be rejected at $\sim\!99\%$ confidence (under the conservative assumption that the uncertainty due to the DM-$z$ is not significantly reduced even with more accumulated data).

\section{Stacking repeating FRBs to improve the bound}\label{sec:Stack}

\vspace{-0.1in}
Before closing, we discuss a possible means of improving the dark matter bound by exploiting repeating FRBs.
Simplistically, for repeating FRBs we can stack their $N$ repetitions, so that the signal gets magnified by a factor of $N$, while the noise only gets magnified by a factor of $\sqrt{N}$, therefore allowing to increase $\overline{R}_f$ by a factor of $\sqrt{N}$ and improve the sensitivity to strong lensing by PBHs. 

This relies on two assumptions: (i) one and all FRB repetitions are lensed by the intervening PBH; (ii) a meaningful fraction of the repetitions have small intrinsic widths so that they can be efficiently stacked together. 

It is straightforward to verify that the first assumption is valid. As long as the time $t_e$ it takes the lensing PBH to cross the effective area given by the angular distance $\theta_E$ in the direction perpendicular to the line of sight~\cite{2017grle.book.....D}\!
\begin{equation}
t_e=\frac{D_L\theta_E}{v_\perp}=\frac{2}{v_\perp}\sqrt{\frac{GM_L}{c^2}\frac{D_{LS}D_L}{D_S}},
\end{equation}
is (much) larger than the time between the first and last FRB repetitions taken into account, we are safe in assuming the same lens for all of them.
Indeed, for a PBH traveling at $v_\perp\!\sim\!300\, {\rm km/s}$, $M_L\!=\!100M_\odot$ and both FRB and PBH at cosmological distances, we get $t_e=\mathcal{O}(100)\,{\rm yrs}$.

As for the second assumption, as we saw in Fig.~\ref{fig:widths hist}, at least for some repeaters there is a large fraction of repeating bursts with small, $\lesssim\!10\,{\rm ms}$, intrinsic widths.

\begin{figure}[h!]
	\includegraphics[width=0.45\textwidth]{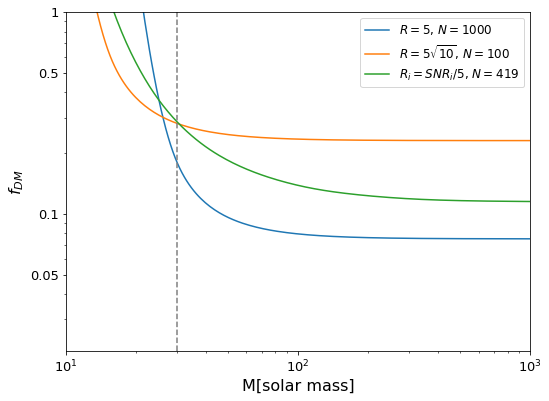}
%	\vspace{-0.15in}
	\caption{Fraction $f_{\rm DM}$ of dark matter allowed in the form of PBHs, calculated for different values of $N_{\rm FRB}$ and an $\overline{R}_f$ matching the number of repetitions, assuming $1000$ non-repeating events (blue), $100$ events with $10$ repetitions each (orange) and the CHIME catalog data with the $N_{\rm FRB}\!=\!419$ events that pass the ${\rm SNR}\geq10$ cutoff and yield a positive redshift (green).  We see that compared to $1000$ non-repeaters, using a factor $10$ fewer events with $10$ repetitions each weakens the bound only a bit overall but at the same time  provides sensitivity to  masses below $M\!=\!30\,M_\odot$ (vertical dashed line).}
	\label{fig:R-N}
\end{figure}

In Fig.~\ref{fig:R-N} we show an example of how stacking repeaters affects the $f_{\rm DM}$ bound. For this example we use Eq.~\eqref{eq:tau_bar} with a constant redshift distribution up to a $z_{cut}\!=\!0.5$ Gaussian cutoff (see e.g.\ Ref.~\cite{Munoz:2016tmg} for details).  
We show the $f_{\rm DM}$ bound for different values of $N_{\rm FRB}$ and an $\overline{R}_f$ matching the number of repetitions. We can see that for $100$ events repeating $10$ times each, stacking  the images  improves upon the results of $1000$ non-repeating events in the lower (LIGO) mass range $ M\!\lesssim\!30\,M_\odot $, where  in the latter case one could barely constrain $f_{\rm DM}$ at all.

\vspace{-0.15in}
%\color{red}
\section{Using the full dynamic spectrum}\label{sec:DS}
%\subsection{Redoing the analysis}

\vspace{-0.05in}

Our analysis hitherto relied on the total light curve of each FRB. Since the flux ratio between the original and lensed images does not depend on the energy (frequency), we could potentially extend this analysis to use the full dynamic spectrum and repeat the analysis using different frequency bins instead of using the total flux~\cite{Paczynski:1987}. As we show below, the SNR of the CHIME FRBs is unfortunately not high enough to allow a search in different bins over the full catalog. However, comparing the results for our lensing candidates  (which are also high-SNR events) in different bins enables a further check of their status.

\begin{figure*}[htbp!]
\centering
\begin{minipage}[b]{0.85\textwidth}
\subfigure{\includegraphics[width=\textwidth]{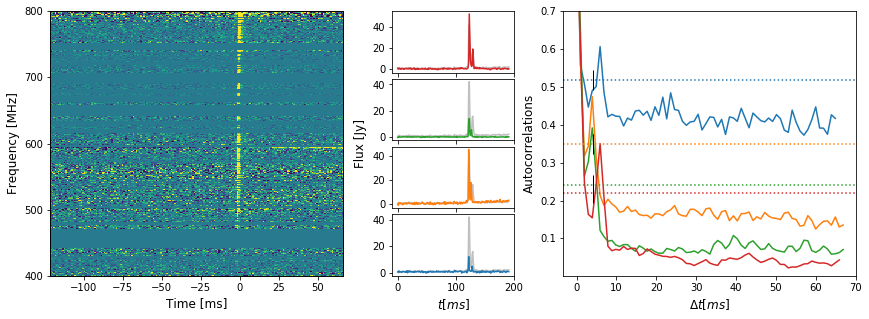}}
\end{minipage}\qquad
\begin{minipage}[b]{0.85\textwidth}
\subfigure{\includegraphics[width=\textwidth]{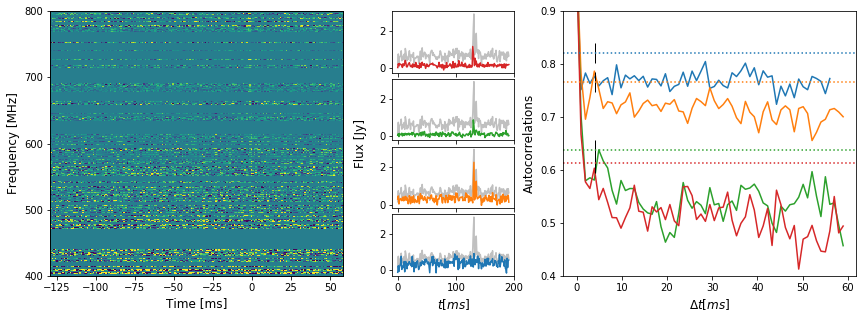}}
\end{minipage}
\begin{minipage}[b]{0.85\textwidth}
\subfigure{\includegraphics[width=\textwidth]{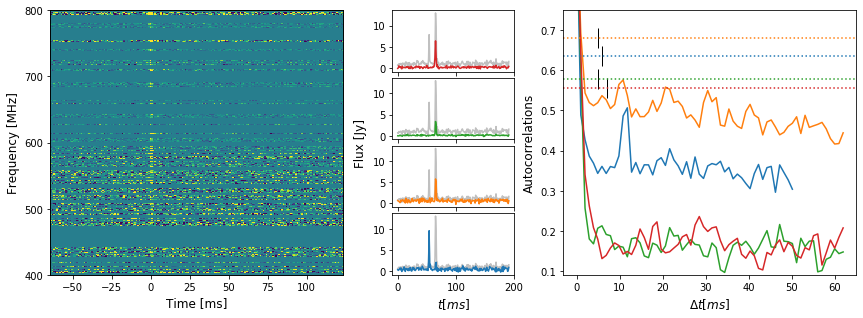}}
\end{minipage}
\begin{minipage}[b]{0.85\textwidth}
\subfigure{\includegraphics[width=\textwidth]{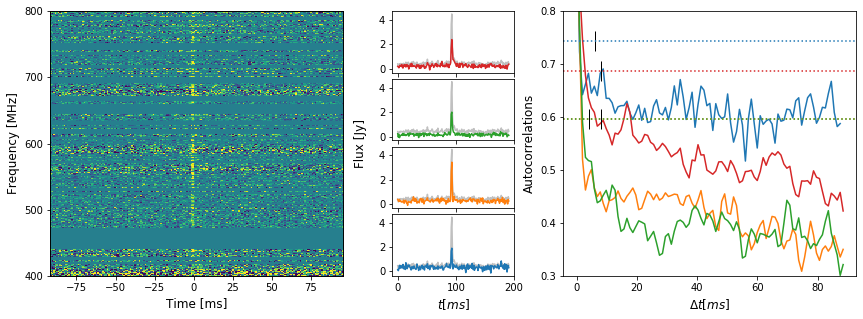}}
\end{minipage}
\caption{%\color{red}
{\it Left panel:} dynamic spectrum. {\it Middle panel:} light curves of the corresponding frequency ranges of the dynamic spectrum in color, with the full spectrum light curve in grey for measure (bottom light curve is for $400-500\rm MHz$ and so on). {\it Right panel:} autocorrelations of the bins' light curves in matching colors, horizontal dotted lines are the thresholds and $\vert$ marks the beginning of each search. {\it From top to bottom:} FRB20190202A, FRB20180916A, FRB20190609B, FRB20190627B.}
\label{fig:Bins}
\end{figure*}

\paragraph{Full search using multiple frequency bins:} We divided the dynamic spectrum (from $400\,{\rm MHz}$ to $800\,{\rm MHz}$) into four equal frequency bins and ran the validation process and detection search separately on each bin. This time we also analyze FRBs wider then $6 \rm{ms}$ as long as the width of the burst in one of the bins is $\leq6$, in order to remain within our target mass range. To filter out low-SNR bins and bins with spurious noise peaks, we only look at bins which have $\rm SNR \geq max(25,SNR_{max}/4)$, where $\rm SNR_{max}$ is the highest $\rm SNR$ of the four bins, and check that the primary peaks in each bin are located within the widths of the primary peak of the full spectrum FRB. The result of the validation procedure on the filtered catalog now leaves us with only 24 FRBs that are valid for a lensing search in more than one bin. This is not nearly enough to constrain $f_{\rm DM}$, so we do not attempt to use these results, but only show a couple of examples in Fig.~\ref{fig:Bins}.  
%The top panel is the validation (with $R=3$) FRB20190202A which passes in all 4 bins, and over the entire spectrum passes with $\rm R=11$. The second panel is the validation (with $R=3$) of FRB20180916A which only passes in two bins and over the entire spectrum passes with $\rm R=4$.
%As previously mentioned, if an FRB is indeed gravitationally lensed, it should show similar autocorrelations regardless of frequency. 
This figure shows the dynamic spectrum, the light curves of the bins and their autocorrelations for several FRBs.
The top panel is the validation (with $R=3$) of FRB20190202A which passes in all 4 bins, and over the entire spectrum passes with $\rm R=11$. The second panel is the validation (with $R=3$) of FRB20180916A which only passes in two bins and over the entire spectrum passes with $\rm R=4$. 

\paragraph{Using a hardness test to scrutinize lensed candidates:}
The third panel in Fig.~\ref{fig:Bins} is the detection search of the third lensing candidate FRB20190609B which fails in all bins. We can see that the pre-burst only exists in the lowest bin where the primary burst is very weak. Therefore we conclude that this candidate is unlikely to be a result of gravitational lensing. The bottom panel is the detection search of the remaining lensing candidate FRB20190627B which also fails the search in all bins.

% +conclusions
\color{black}

\section{Conclusions}\label{sec:conclusions}
In this work we have extended previous forecasts on the limit that can be placed on PBH dark matter with strong lensing of FRBs by setting the bound based on the actual burst widths and on individually calibrated detectability thresholds for each event. The dedicated validating procedure we used to set the flux-ratio thresholds for detectability relied on simulating lensing echos added to each event and running a detection algorithm (based on autocorrelation) to identify them.  We also accounted for the uncertainty in the relation between the FRB dispersion measure and source redshift, which with current knowledge of the FRB-to-redshift relation, adds significant error bars to the currently inferred threshold.

We found that in data from the first CHIME catalog, only 114 out of 536 events pass the validation process and can be used reliably for the lensing search. Unfortunately, this does not suffice for producing a meaningful bound on the fraction of dark matter in PBHs, but with thrice the current data, the bound will start to become prohibitive. We also tried removing the SNR filtering of the FRBs, which left us with 353 FRBs to be validated. To filter out spurious detections we had to increase to $2.8\sigma + \mu$, which in turn lowered $R_f$ for some of the events so that with 145 valid FRBs we got a slightly weaker bound than the one shown in Section ~\ref{sec:results}. 

Interestingly, regardless of the SNR filtering, our detection algorithm was steadfast in identifying one candidate lensing event, FRB20190627B, which deserves follow-up at higher time resolution, ideally with proper Bayesian analysis~\cite{Paynter:2021wmb,Yang:2021wwd,Veres:2021gfr}, to corroborate, or more likely refute, the lensing hypothesis to explain the double burst. 

We also showed that for low masses ($ M \lesssim 30\,M_\odot$), increasing sensitivity contributes more to the $f_{\rm DM}$ bound than increasing the number of FRBs. We suggested to do so by stacking repeating FRBs (a focus of several recent strong lensing analyses~\cite{Li:2017mek,Zitrin:2018let,Wucknitz:2020spz,Abadi:2021ysz}) and demonstrated the potential to improve the constraints on PBH dark matter. 

%{\color{red}
Finally, we discuss the potential to use the full dynamic spectrum to improve the sensitivity and perform a frequency/energy hardness test on our lensing candidates. 
Unfortunately, our validation procedure shows that the SNR of current CHIME data is not enough to allow a full lensing search in individual bins. 
Nevertheless, examining the autocorrelation of our lensing candidates in each bin casts doubt on FRB20190609B as a lensed candidate.%we where not able to reach a definitive conclusion regarding FRB20190627B.
%}

While strong lensing of FRBs may not be competitive with other methods to probe  $ M\! \gtrsim\! 10\,M_\odot$ PBH dark matter in the very near future, it ultimately presents a unique means of probing this scenario. This work provides strong motivation for the study of higher-resolution data (compared to what is publicly available) to extract a PBH dark matter bound. In principle, the extraordinarily-high-resolution time series obtainable by CHIME (down to several $\mu {\rm s}$~\cite{CHIMEFRB:2018mlh,Pleunis:2021qow}, using coherent dedispersion~\cite{Hankins} of the recorded electric field~\cite{Michilli:2020wbi}) could allow to detect overlapping bursts as a result of strong lensing by much lower masses, but the effect of scintillation due to the intervening  medium on the source and lensed images~\cite{Macquart:2018rsa,Schoen:2021xhp} will inevitably limit this effort~\cite{Eichler:2017eid,Katz:2019qug}.

{\it Note added:} While the results of our analysis were being drafted into the first version of this manuscript, Ref.~\cite{Zhou:2021ndx} appeared, which also carries out an analysis of CHIME data to search for lensing candidates. Our approach differs in several respects. First and foremost, we do not adopt a single flux-ratio threshold for all events when calculating the bound on PBH dark matter nor a single SNR-based criterion, but rather calibrate one for each of the FRBs in the catalog based on our validation procedure, which reduces the number of FRBs that can be used to derive the bound by a factor of $\sim$10. In addition, we include the uncertainty from the DM-$z$ relation and advocate for using repeating FRBs to improve the bound on lower masses in the future. Interestingly, while one of the candidates we identified in our search, FRB20181104C, appears in their list of lensing candidates (and is subsequetly ruled out due to dynamic spectrum differences between the two peaks), our surviving candidate, FRB20190627B, does not show on their list.
 
%but this will be limited by the effects of scintillation due to the intervening  medium on the source and lensed images

\section*{acknowledgements}

We thank Tal Abadi for useful discussions, Julian Mu\~noz and Lingyuan Ji for helpful feedback on the manuscript and Yuval Amrami for technical assistance.  { We also thank the anonymous referee for useful suggestions which helped improve the quality of the paper.} EDK acknowledges support from an early-career faculty fellowship awarded by the Azrieli Foundation.

%------------------------------------------------------------------------------

%------------------------------------------------------------------------------

\end{document}